\documentclass{elsart}
\usepackage{amssymb}
\newcommand{\sect}[1]{\setcounter{equation}{0}\section{#1}}

\usepackage{latexsym}
\usepackage{graphics}
\begin{document}
\begin{frontmatter}
\title{A mechanism to derive multi-power law functions: an application in the
econophysics framework.}

\author{A.M. Scarfone}

\address{Istituto
Nazionale di Fisica della Materia (CNR-INFM) and Physics Department\\
Unit\'a del Politecnico di Torino, Corso Duca degli Abruzzi 24,\\
I-10129 Torino, Italy}

\date{\today}
\begin {abstract}
It is generally recognized that economical systems, and more in
general complex systems, are characterized by power law
distributions. Sometime, these distributions show a changing of the
slope in the tail so that, more appropriately, they show a
multi-power law behavior. We present a method to derive analytically
a two-power law distribution starting from a single power law
function recently obtained, in the frameworks of the generalized
statistical mechanics based on the Sharma-Taneja-Mittal information
measure. In order to test the method, we fit the cumulative
distribution of personal income and gross domestic production of
several countries, obtaining a good agreement for a wide range of
data.
\end {abstract}

\end{frontmatter}
\maketitle

\sect{Introduction}

Free-scale behavior in the economical systems have been observed
since 19th century, when Pareto noticed that the cumulative
distribution of the personal income $P(x)=\int_x^\infty p(y)\,dy$ of
several countries behaves like a power law function. Afterwards,
Gibrat clarified that such a power law behavior holds only for the
high income region, whilst in the low-middle income region, which
includes almost the whole body of data, the curve is well fitted by
a log-normal
distribution. \\
Actually, the problem concerning the real profile showed by the
function $P(x)$ in the whole range of the accessible data is still
an open question. In particular, it has been suggested
\cite{Gallegati} that deformed exponential functions derived
recently in the field of the generalized statistical mechanics, can
be fruitfully employed to modeling analytically the cumulative
distribution $P(x)$ for a wide range of the income values.\\
Notwithstanding, the recent analysis based on a huge quantity of
data nowadays accessible, shown that sometime the crossover among
the low-middle region (the log-normal region) and the high region in
the upper tail of the distribution (the Pareto region, with a power
law behavior $P(x)\sim x^{-s}$, where $s$ is a positive constants
quite generally $1\leq s\leq2$), does not occur smoothly, giving
origin to {\em knee or ankle effects} (see for instance
\cite{Yakovenko}). Moreover, in some cases, it has been observed a
deviation from the Pareto behavior in the highest region, which can
originate a new power law behavior $P(x)\sim
x^{-\tilde s}$ with a different slope $\tilde s\not=s$.\\
The complicate profile in the shape of $P(x)$ cannot be accounted
for by a generalized exponential with a single power law behavior.
This open the questions: how can we describe the shape observed in
$P(x)$ with an analytically simple function? \\
In the present contribution, we introduce a mechanism which permits
to generate multi-power law functions by employing deformed
exponentials
and logarithms with a single power law asymptotic behavior.\\
Notice that, two-power law behavior have been observed in various
economical systems like, for instance, in the cumulative
distribution of the personal income \cite{Borges}, in the cumulative
distribution of the land price \cite{Ishikawa} or in the returns of
many market indexes \cite{Gopikrishnan}.\\ On a general ground,
two-power law behavior have been observed in different physics
fields as well as in biological, geological and social sciences.
Among the many, we quote the dielectric relaxation \cite{Weron}, the
re-association in folder proteins \cite{Tsallis0}, and others
\cite{Montemurro}.\\ It is worthy to remark that there have been
proposed different methods in literature
\cite{Tsallis0,Naudts1,Wang} to produce generalized distributions
with a double-power law behavior which differ from the one advanced
in the following.

\sect{Deformed logarithms and exponentials} Generalized exponential
functions ${\mathcal E}(x)$, interpolating between the standard
exponential $\exp(x)$ for $x\ll1$ and the power law $x^{-s}$ for
$x\gg1$, arise naturally in the study of thermostatistic proprieties
of complex systems which show free-scale feature. In
\cite{Scarfone1}, it has been postulated a very general expression
for the entropy of such a system
\begin{equation}
S(p)=-\int p(x)\,\Lambda(p(x))\,dx \ ,\label{ent}
\end{equation}
(in the unity of Boltzmann constant $k_{\rm B}=1$), where
$\Lambda(x)$ plays the role of a generalized logarithm, the inverse
function of ${\mathcal E}(x)$. By requiring that the distribution,
derivable through a variational problem, assumes the form
\begin{equation}
p(x)={\mathcal E} \left(-\sum_{j=1}^M\beta_j\,|x|^{\mu_j}\right) \
,\label{pdf}
\end{equation}
which mimics the well-known Boltzmann-Gibbs distribution, the
following functional equation has been obtained
\begin{equation}
\frac{d}{dx}\left[x\,\Lambda(x)\right]=\lambda\,\Lambda\left(\frac{x}{\alpha}\right)
\ .\label{fde}
\end{equation}
Here, $\alpha$ and $\lambda$ are constants given by
\begin{equation}
\alpha=\left|\frac{1+r-\kappa}{1+r+\kappa}\right|^{1/2\,\kappa} \
,\hspace{10mm}\lambda=\frac{|1+r-\kappa|^{(r+\kappa)/2\,\kappa}
}{|1+r+\kappa|^{(r-\kappa)/2\,\kappa}} \ .\label{al}
\end{equation}
The quantities $\beta_j$ in Eq. (\ref{pdf}) play the role of
Lagrange multipliers associated to the $M$ constraints
$\int|x|^{\mu_j}\,p(x)\,dx={\mathcal O}_j$ which represent the
$\mu_j$-th momenta of $x$. Typically, the
 constants $\mu_j$ are integers (for instance, $\mu_1=0$
gives the normalization $\int p(x)\,dx={\mathcal O}_1$, $\mu_2=1$ is
the mean value $\langle x\rangle={\mathcal O}_2$, and so on) but for
sake of generality we assume $\mu_j\in I\!\!R$.\\
The most general solution of Eq. (\ref{fde}), accounting for the
boundary conditions $\Lambda(1)=0$ and
$(d/dx)\,\Lambda(x)\Big|_{x=1}=1$, derived from certain physically
and mathematically justified assumptions, is given by
\begin{equation}
\Lambda(x)\equiv\ln_{_{\{\kappa,\,r\}}}(x)=x^r\,\frac{x^\kappa-x^{-\kappa}}{2\,\kappa}
\ ,\label{log}
\end{equation}
which recover the standard logarithm in the $(\kappa,\,r)\to(0,\,0)$
limit.\\By requiring that $\ln_{_{\{\kappa,\,r\}}}(x)$ is a
continuous, monotonic, normalizable, concave and increasing function
for $x\in(0,\,+\infty)$, we obtain the restrictions $-|\kappa|\leq
r\leq|\kappa|,$ if $0\leq|\kappa|<1/2$ and $|\kappa|-1\leq
r\leq1-|\kappa|,$ if $1/2\leq|\kappa|<1$. Notwithstanding, for
particular applications some of the above mathematical requirements
can be relaxed permitting less restrictive conditions for the
deformation parameters. For instance, in certain practical
situations one is welling with a normalization in a finite interval
$x\in(0,\,x_{\rm max})$ \cite{Newman} and we can discard the condition $|\kappa|<1$.\\
In the following, we require only that Eq. (\ref{log}) be a
monotonic function, so that its inverse function, the generalized
exponential $\exp_{\{\kappa,\,r\}}(x)$, certainly exists. This is
accomplished by requiring only that $-|\kappa|<r<|\kappa|$.
\\
From Eq. (\ref{log}) we obtain that $\ln_{_{\{\kappa,\,r\}}}(x)\to
x^{r+|\kappa|}/|2\,\kappa|$ for $x\to+\infty$ and
$\ln_{_{\{\kappa,\,r\}}}(x)\to -x^{r-|\kappa|}/|2\,\kappa|$ for
$x\to0$, whilst $\ln_{_{\{\kappa,\,r\}}}(x)\to (x-1)$ for
$|x-1|\ll1$. In the same way, we have that
$\exp_{\{\kappa,\,r\}}(x)\to|2\,\kappa\,x|^{1/(r\pm|\kappa|)}$ for
$x\to\pm\infty$, whilst $\exp_{\{\kappa,\,r\}}(x)\to1+x$ for
$x\to0$. Thus, the deformed exponential $\exp_{\{\kappa,\,r\}}(x)$
interpolates with continuity between the standard exponential
$\exp(x)\simeq1+x$, for
$x\to0$, and the power law $|x|^{-s}$ with slope $s=-1/(r\pm|\kappa|)$, for $x\to\pm\infty$. \\
Finally, accounting for the solution (\ref{log}), the entropy
(\ref{ent}) assumes the form
\begin{equation}
S_{_{\kappa,\,r}}(p)=-\int
p(x)\,\ln_{_{\{\kappa,\,r\}}}\Big(p(x)\Big)\,dx \ ,\label{stm}
\end{equation}
which recovers, in the limit $(\kappa,\,r)\to(0,\,0)$, the
Shannon-Boltzmann-Gibbs entropy $S=-\int p(x)\,\ln p(x)\,dx$. This
entropic form, introduced previously in literature in
\cite{Taneja1,Mittal1,Mittal2}, is known as the Sharma-Taneja-Mittal
information measure and has been applied recently in the formulation
of a possible thermostatistics theory \cite{Scarfone3,Scarfone4}.

\sect{Two-power law function}

Endowed with the deformed logarithm $\ln_{_{\{\kappa,\,r\}}}(x)$ and
the deformed exponential $\exp_{_{\{\kappa,\,r\}}}(x)$ we can
construct the quantity
\begin{equation}
\Pi_{\sigma_1}(x)=\exp_{_{\{\kappa_1,\,r_1\}}}\left(a_1\,\ln_{_{\{\kappa_1,\,r_1\}}}(x)\right)
\ ,\label{p}
\end{equation}
where $\sigma_1$ denotes the set of parameters
$\sigma_1\equiv(\kappa_1,\,r_1,\,a_1)$, with $a_1\geq1$. The
function (\ref{p}) is therefore employed in the following
construction
\begin{equation}
f(x)=\Pi_{\sigma_1}\circ\exp_{_{\{\kappa_2,\,r_2\}}}(x)\equiv
\Pi_{\sigma_1}\left(\exp_{_{\{\kappa_2,\,r_2\}}}(-x)\right)\
.\label{double}
\end{equation}
We observe that, for $a_1=1$ expression (\ref{double}) reduces to
$\exp_{_{\{\kappa_2,\,r_2\}}}(-x)$, for $(\kappa_1,\,r_1)\to(0,\,0)$
we obtain $[\exp_{_{\{\kappa_2,\,r_2\}}}(-x)]^{a_1}$, whilst for
$(\kappa_1,\,r_1)=(\kappa_2,\,r_2)$ we obtain
$\exp_{_{\{\kappa_1,\,r_1\}}}(-a_1\,x)$.\\
Accounting for the asymptotic behavior of the deformed exponential
and logarithm we can distinguish three regions in the range $x>0$ of
$f(x)$. A first region, for $a_1\,x\ll1$, characterized by the
linear behavior
\begin{equation}
f(x)\sim1-a_1\,x \ ,
\end{equation}
like the exponential $\exp(-a_1\,x)$ does for $x\to0$. A second
intermediate region, for $x\ll1\ll a_1\,x$, where $f(x)$ is
characterized by the power law behavior
\begin{equation}
f(x)\sim x^{-s_1} \ ,
\end{equation}
with slope $s_1=1/(|\kappa_1|-r_1)$.\\ Finally, for $x\gg1$ we
obtain the asymptotic power law behavior
\begin{equation}
f(x)\sim x^{-s_2} \ ,
\end{equation}
whose slope is now $s_2=1/(|\kappa_2|-r_2)$.\\
Thus, $f(x)$ behaves like a power law function both in the middle
and in the far region of $x>0$ with slopes $s_1$ and $s_2$,
respectively. In
this sense, we call Eq. (\ref{double}) a two-power law function.\\
Form the above analysis, we easily realize that the constant $a_1$,
introduced in the definition of $\Pi_{\sigma_1}(x)$, gives
approximatively the width
of the intermediate region having slope $s_1$.\\
\begin{figure}[ht]
\begin{center}
 \resizebox{140mm}{!}{\includegraphics{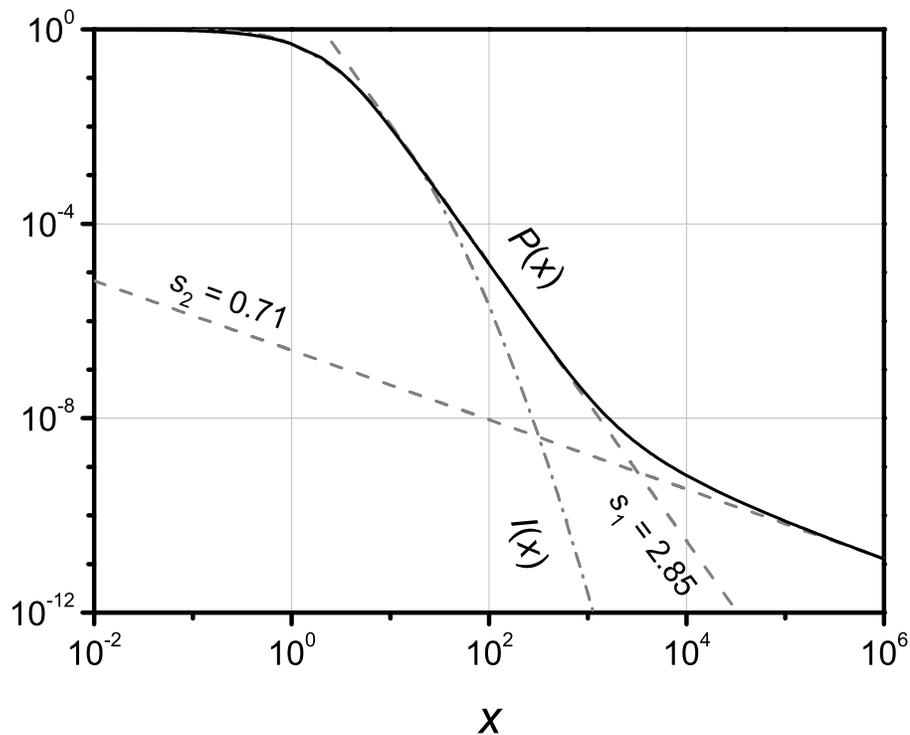}} \vspace{-5mm}
\caption{Log-log plot of the two-power law function (\ref{double})
(solid line). The dot-dashed line is the cumulative integral of the
log-normal function. The dashed lines denote the asymptotic
extension of the function $P(x)$ in the two-power law regions.}
\end{center}
\end{figure}
As an example, let us specialize Eq. (\ref{double}) to the case
$r=0$. In this situation, the generalized exponential and logarithm
assume, respectively, the expression \cite{Kaniadakis}
\begin{equation}
\exp_{\{\kappa\}}(x)=\left(\kappa\,x+\sqrt{1+\kappa^2\,x^2}\right)^{1/\kappa}
\ ,
\end{equation}
and
\begin{equation}
\ln_{\{\kappa\}}(x)=\frac{x^\kappa-x^{-\kappa}}{2\,\kappa} \ .
\end{equation}
In figure 1, we plot the function
\begin{equation}
P(x)=\exp_{\{\kappa_1\}}\left(a_1\,\ln_{\{\kappa_1\}}\left(\exp_{\{\kappa_2\}}
\left(-x\right)\right)\right) \ ,\label{doub}
\end{equation}
for the values $\kappa_1=0.35$, $\kappa_2=1.4$ and $a_1=10^3$. In
the same graphic, the dot-dashed line depicts the cumulative
integral $I(x)=\int_x^\infty p(y)\,dy$ of the log-normal
distribution
\begin{equation}
p(x)={1\over(2\,\pi)^{1/2}\,x}\exp\left(-{1\over2}\,\ln^2x\right) \
.
\end{equation}
The dashed lines represent the asymptotic prolongation of the power
law behavior of $P(x)$ whose slopes are given, respectively, by
$s_1=1/k_1=2.85$ and $s_2=1/k_2=0.71$. We observe a good agreement
between the functions $I(x)$ and $P(x)$ only in the low region of
$x$.

\sect{Application to econophysics}

In the following, we employ the function derived in the previous
Section 3 to fit some distributions data
obtained in the economy framework.\\
We pose $P(x)=f(-\beta\,|x|^\mu)$, with $\beta$ and $\mu$ fitting
parameters, the cumulative distribution representing the
probability of finding a value $X$ equal to, or greater than $x$.\\
In figure 2, we present the results of the fit (in log-log scale)
for the data of the inverse cumulative distribution of the personal
income of Japan (1975) obtained in \cite{Souma} and  USA (2000)
obtained in \cite{Borges}, as well as, the data of the inverse
cumulative distribution of the gross domestic production of Brazil
(1996) and Germany (1998) obtained in \cite{Borges}.\\ In every
graphic, we report the dashed lines representing the asymptotic
behavior of $P(x)$ in the two power law regions with slope given by
$s_1=\mu/k_1$ and $s_2=\mu/k_2$, respectively.\\
\begin{figure}[t]
\begin{center}
  \resizebox{140mm}{!}{\includegraphics{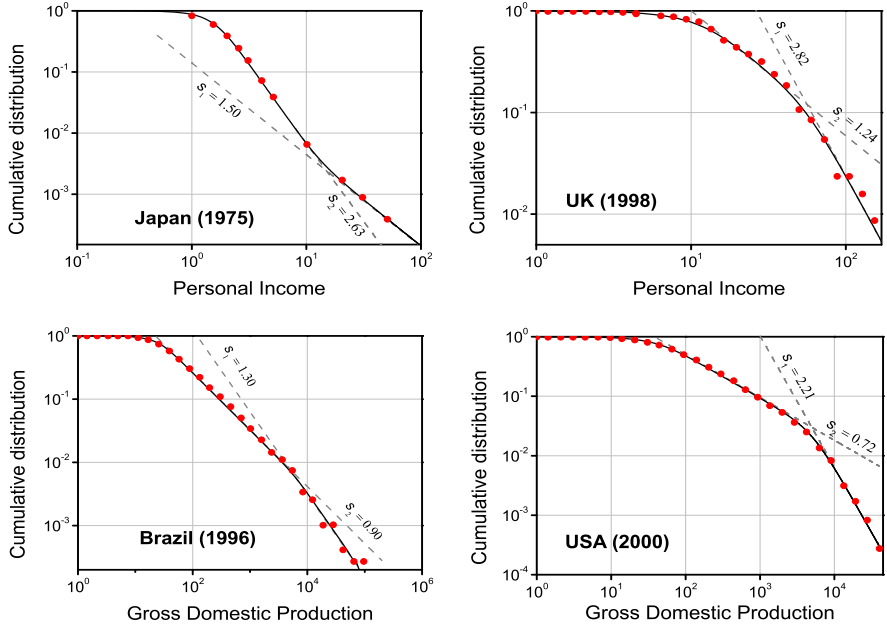}} \vspace{-5mm}
\caption{Log-log plot of personal income distribution for Japan
(1975) \cite{Souma} and USA (2000) \cite{Borges} and gross domestic
production distribution for Brazil (1996) and Germany (1998)
\cite{Borges}. The solid line represents the fit obtained with the
two-power law function (\ref{doub}). The straight dashed lines are
plotted for convenience to indicate the asymptotic power-law
prolongation.}
\end{center}
\end{figure}
The data fit are reported in table 1.
\begin{center}
Table 1.\\
Parameters for the cumulative distribution
$P(x)$.\\
\begin{tabular}{lccccc}\hline
Country & $\kappa_1$ & $\kappa_2$ & $a_1$ & $\mu$ & $\beta$\\
\hline
Japan (1975) & 1.14 & 2.00 & 390 & 3.00 & $3.5\cdot10^{-4}$ \\
UK (1998) & 1.70 & 0.75 & 8 & 2.12 & $2.5\cdot10^{-4}$ \\
Brazil (1996) & 2.20 & 1.53 & $2\cdot10^4$ & 1.99 & $2.3\cdot10^{-8}$ \\
USA (2000) & 2.00 & 0.65 & 231 & 1.44 & $6.0\cdot10^{-6}$ \\
\hline
\end{tabular}
\end{center}
The crossover between the first and the second power law region,
causing a reduction of the slope, with $s_2<s_1$ (UK, Brazil and
USA), is named {\em kink effect} \cite{Borges}. Similarly, the
crossover between the first and the second power law region causing
an increase of the slope, with $s_2>s_1$ (Japan), is named {\em
ankle effect}.

\sect{Generalization} Let us briefly discuss the generalization of
the method introduced in Section 3 in order to generate functions
with more than two power law behavior. This can be accomplished
starting from the building block function
\begin{equation}
\Pi_{\{\sigma_i\}}(x)=\exp_{\{\kappa_i,\,r_i\}}\left(a_i\,\ln_{\{\kappa_i,\,r_i\}}(x)\right)
\ ,
\end{equation}
and introducing the quantity
\begin{equation}
\Pi_{\{\vec\sigma\}}(x)=\Pi_{\{\sigma_1\}}\circ\Pi_{\{\sigma_2\}}
\circ\ldots\circ\Pi_{\{\sigma_{n-1}\}}(x) \ ,
\end{equation}
\begin{figure}[ht]
\begin{center}
  \resizebox{140mm}{!}{\includegraphics{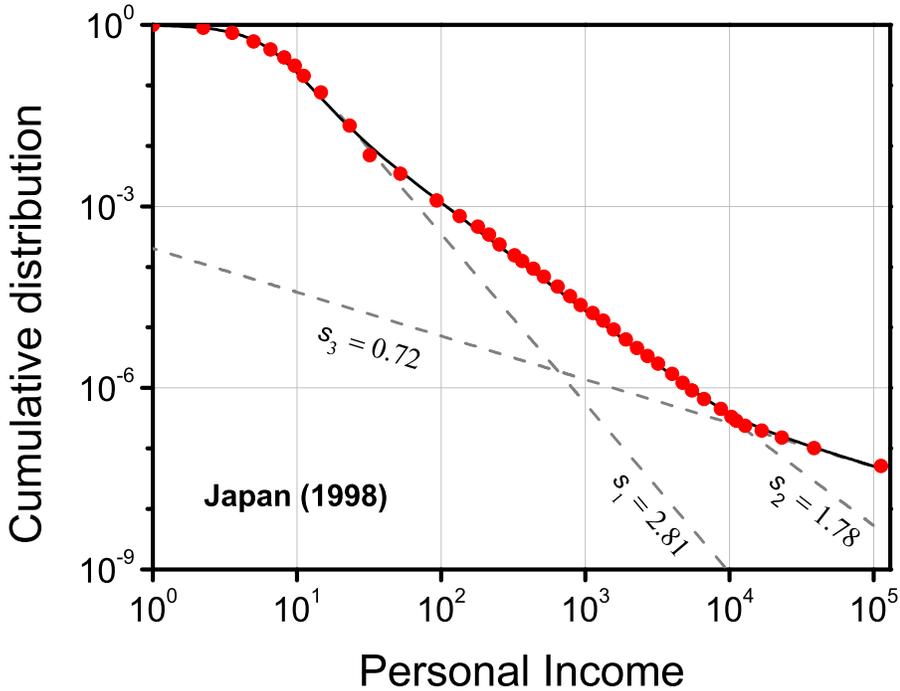}} \vspace{-5mm}
\caption{Log-log plot of the cumulative distribution of Japan for
the year 1988. It is observed a deviation from the Pareto behavior
in the highes income region ($x>10^4$). }
   \end{center}
\end{figure}
where
$\vec\sigma\equiv(\sigma_1,\,\sigma_2,\,\ldots,\,\sigma_{n-1})$ is a
$(n-1)$-vector whose $i$th entry $\sigma_i=(\kappa_i,\,r_i,\,a_i)$
contains the relevant informations about the slope and the width of
the $i$th power law region. It is easy to verify that the function
\begin{equation}
f(x)=\Pi_{\{\vec\sigma\}}\circ\exp_{\{\kappa_n,\,r_n\}}(-x) \
,\label{multi}
\end{equation}
exhibits a $n$-power law behavior.\\
In figure 3, we report the fit of the 1998 Japanese income data
obtained in \cite{Souma1} by employing the function
\begin{equation}
P(x)=\Pi_{\{\vec\sigma\}}\circ\exp_{\{\kappa_3\}}(-\beta\,x^\mu) \ ,
\end{equation}
derived from Eq. (\ref{multi}) for $n=3$ and $r_i=0$. The fitting
data are
$\kappa_1=0.71,\,\kappa_2=1.12,\,\kappa_3=2.77,\,a_1=10,\,a_2=4\cdot10^5$,
$\beta=6.00\cdot10^{-9}$ and $\mu=2.00$.

\sect{Conclusions} We have derived a simple method which permits to
generate functions with a multi-power law behavior starting the
deformed logarithm $\ln_{\{\kappa,\,r\}}(x)$ and the deformed
exponential $\exp_{\{\kappa,\,r\}}(x)$, recently derived in
\cite{Scarfone1}, which exhibit a single power law profile. An
explicit two-power law function has been constructed starting from
the $\kappa$-exponential and its inverse, the $\kappa$-logarithm. We
have employed this function to fit the inverse cumulative
distribution of the personal income and of the gross domestic
production of several countries, showing a good agreement among the
analytical and the empirical data for a wide range of values.

\end{document}